\documentclass[aps,prb,12pt,notitlepage]{revtex4-1}

\usepackage[T1]{fontenc}
\usepackage[latin1]{inputenc}
\usepackage{amsmath}
\usepackage{graphicx}
\usepackage{amssymb}

\usepackage{color} 
\usepackage{graphicx}
\usepackage{bm}
\usepackage{amsmath}

\usepackage{gt15}

\begin{document}
\title{
Green's function representation of spin pumping effect
}
\author{Gen Tatara}
\affiliation{RIKEN Center for Emergent Matter Science (CEMS), 
2-1 Hirosawa, Wako, Saitama, 351-0198 Japan}

\date{\today}

\begin{abstract}
In this study,  current pumping by an external potential is studied on the basis of the Keldysh Green's function method, and 
a pumping formula written in terms of retarded and advanced Green's functions is obtained.
It is shown that pumping is essentially driven by a change of particle distribution before and after an external perturbation. 
The formula is used to study the spin pumping effect in the case of strong \sd exchange interaction, and the driving field is identified to be the spin gauge field.
At the lowest order in the precession frequency of magnetization, the spin gauge field works as a constant potential, and the system is shown to reduce to a static problem of spin current generation by a time-independent potential with off-diagonal spin components.
\end{abstract}  

\maketitle

\newcommand{\hop}{{t}}
\newcommand{\jzero}{j^{(0)}}
\newcommand{\jzerov}{\jv^{(0)}}
\newcommand{\jlam}{j^{\lambda}}
\renewcommand{\ctil}{\widetilde{c}}
\renewcommand{\dtil}{\widetilde{d}}
\newcommand{\rhov}{\bm{\rho}}
\newcommand{\els}{\ell_{\rm s}}

\section{Introduction}

The spin pumping effect, in which a spin current is induced by the  precession of magnetization in a ferromagnetic normal metal junction, is the basis of some of the most important technologies for spin current generation.
The effect was theoretically derived by Silsbee \cite{Silsbee79}, who noted that when a microwave is applied to a  junction, a dynamic spin accumulation is generated at the interface, which in turn leads to a flow of spin (spin current) as a result of electron diffusion.  
A similar type of spin current generation was later discussed by Tserkovynak et al. in 2002 \cite{Tserkovnyak02} based on an application of scattering theory for adiabatic pumping effects; this effect is known as the spin pumping effect \cite{Tserkovnyak02b}.

Studies of adiabatic pumping were initiated by a seminal 1983 paper by Thouless \cite{Thouless83}, in which  transport induced by an adiabatic change of a potential was discussed in light of Berry's phase attached to the wave function.  
Current generation by a slowly oscillating potential was later discussed by B\"uttiker et al. in the context of scattering matrix theory \cite{Buttiker94}.
The scattering theory of adiabatic pumping was generalized in the case of periodic variation by Brouwer \cite{Brouwer98}, who presented  a formula describing the pumped charge by means of a scattering matrix. 
It was also pointed out that the variation of two independent parameters is necessary for adiabatic pumping.
Tserkovynak et al. applied  Brouwer's formula to a case of a junction between a ferromagnet and a normal metal  and argued that a spin current is generated when the magnetization is dynamic. 
Tserkovnyak et al. further noted that the amount of the pumped spin current is governed by what they termed a spin mixing conductance, which is related to a spin-dependent transmission coefficient.
The dynamics of a magnetic system can be described by two parameters (the $x$ and $y$ components for the case of precession around the $z$ axis), and thus the application of an adiabatic pumping theory requiring two independent driving parameters is reasonable.

The spin pumping effect in Ref. \cite{Tserkovnyak02} was originally discussed to explain the enhancement of magnetic damping discovered experimentally by Mizukami et al \cite{Mizukami01}.
As the effect turned out to be a convenient experimental means for injecting spin current into metals, their work produced much stimulus for further experimental work, and the effect has been employed in a large number of recent studies \cite{Maekawa13}. 
Spin pumping theory was successful in the sense that it explained the effect by introducing a new phenomenological parameter - spin mixing conductance-  and provided a convenient tool for interpreting experimental results.
The spin mixing conductance can be estimated by  first-principles calculations combined with a calculation of scattering properties \cite{Xia02,Jia11}.  
However, as was pointed out recently in Ref.  \cite{Chen15}, the formalism has a disadvantage from the viewpoint of material design, as the estimation of spin pumping efficiency requires solving a scattering problem and therefore the efficiency is not directly related to the material parameters.
Furthermore, as the theory was developed by borrowing the formalism for adiabatic pumping, and because it 
is only concerned with the spin mixing conductance, the physical mechanism of spin pumping remains obscure. 
Without addressing the nature of the driving force or field for the spin current generation,  
scattering theory may be able to provide an explanation of the phenomenon 
in which  pumping is caused by a time variation of magnetization that modifies the scattering potential. 
In this paper, we revisit the spin pumping effect in order to clarify its physical mechanism. 
By identifying the driving field, the relation between spin pumping and adiabatic pumping is expected to become clearer.

The issues of conventional spin pumping theory were studied and analternative formulation was presented in a recent paper by Chen and Zhang \cite{Chen15}.
They considered a small-amplitude oscillation of magnetization for which they calculated the spin current without using scattering theory and obtained an expression for the spin-pumping coefficient in terms of retarded and advanced Green's functions.
Unlike the scattering approach, their formalism is applicable to systems with disorder and spin relaxation and it was used to study the effect of interface spin-orbit interaction.
Although the difficulties describing spin pumping in terms of scattering were resolved by Chen and Zhang, there remain issues to be further investigated, including how to clarify the fundamental difference and/or similarity between spin pumping and adiabatic charge pumping.
Although a general pumping formula based on Berry's phase formulation \cite{Brouwer98} to address adiabatic pumping is mathematically elegant, it is not practical for material designs aiming at spintronics applications.

The objective of this  paper is to develop a practical formula for pumping effects and to use it as a basis for discussing the spin pumping effect.
In the case of electrons driven by a general  external potential, we derive a linear response formula for the pumping effect that is written in terms of retarded and advanced Green's functions and therefore can be smoothly incorporated into a first-principles calculation scheme. 
The relation of this formulation to the scattering approach was discussed within the argument put forth by Fisher and Lee \cite{Fisher81}.
We show generally that the origin of the pumping is a non-commutativity between  the external perturbation and the operator representing the particle distribution.
In the dynamic-potential case, the non-commutativity arises because the particle energies before and after the application of the dynamic perturbation are different, as argued in Ref. \cite{Buttiker94}. 
In the present Green's function representation, the topological meaning of adiabatic pumping, discussed based on scattering matrix formulation \cite{Brouwer98,Moskalets02}, appears not clearly seen.

The case of spin pumping is studied in detail by use of a unitary transformation in the spin space in order to correctly grasp  the low-energy  properties \cite{TKS_PR08}. 
Thus our results are not restricted to a small-amplitude case discussed in Ref. \cite{Chen15}.
The driving field for spin pumping is identified as the  non-adiabatic components of the spin gauge field $A_{\rm s}$, which is the linear order in the time derivative of the direction of the magnetization.
At the linear order in the precession frequency, the spin gauge field containing the first-order time derivative is treated as static; it therefore works as a static potential that causes spin mixing. 
It is shown that, even though the problem reduces to a static one,  a spin current is pumped as a result of the non-commutativity of the spin-dependent potential (spin gauge field). 
The mathematical mechanism for spin pumping viewed in the rotated frame as a response to the driving potential (spin gauge field) is therefore different from that in adiabatic charge pumping, in which the dynamic nature of the driving potential is essential, although it is obvious that spin pumping is a dynamic effect driven by dynamic magnetization.

The fact that the spin gauge field is the physical field of spin pumping is consistent with experimental observations that the spin accumulation at the interface of a ferromagnet and a normal metal is greatly enhanced when magnetization becomes dynamic, while only a tiny accumulation is present in the static case \cite{Hou16}. 
Thus, the non-equilibrium spin accumulation mechanism driven by the spin gauge field is far more efficient than the one caused by a static magnetic-proximity effect. 

The physics of the spin pumping effect can be studied using a simplified  model of a dynamic magnetic dot coupled to electron reservers or leads \cite{Wang03,Wang04,Splettstoesser08,Rojek14}.
Phase shift argument was presented in Ref. \cite{Zhou03}.
Recently, a full counting statistics approach was employed to study the spin current from a dot under a magnetic field \cite{Nakajima15}.

\section{Potential model for pumping}
\begin{figure}
	\includegraphics[width=0.4\textwidth]{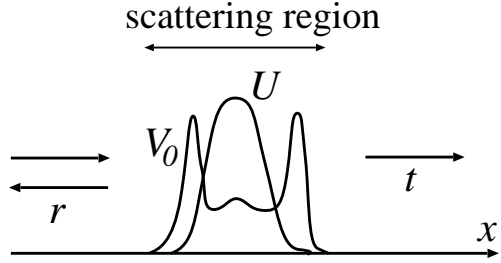}
	\caption{ Schematic depicting the system under consideration. A static potential $V_0$ is localized in  the scattering region. We evaluate charge and spin currents in the asymptotic region when an external driving potential $U$ is applied in the scattering region.  
	Pumping effects in a junction are simulated by taking into account the barrier potentials at the interfaces in the static potential $V_0$.  
	}
	\label{FIGpot}
\end{figure}

In this section we consider a pumping effect in a system of electrons in a potential $V\equiv V_0+U$, where $V_0$ is a static potential and $U$ is a driving potential for the current.
We consider the static potential $V_0$ to have solely diagonal components in spin, while  the driving potential $U$ may have off-diagonal components and may be dynamic.
Both potentials are localized in the scattering region (Fig. \ref{FIGpot}).
The field representation of the Hamiltonian can be given in terms of the double-component field  operators $c$ and $c^\dagger$  ($c=(c_+,c_-)$ with $\pm$ being the spin indices as
\begin{align}
  H &= \sum_{\rv} 
 c^\dagger \lt(-\frac{\nabla^2}{2m} +V_0(\rv)+U(\rv,t)\rt) c.
\label{Hgeneral}
\end{align}
Although the Hamiltonian (\ref{Hgeneral}) can be solved quantum mechanically, we use a field representation and the Green's function method, as transport properties such as current generation can be analyzed in a straightforward manner. 
The disadvantage of the quantum-mechanical approach is that the information on particle occupation (the Fermi distribution function) is not straightforwardly incorporated.
Once a linear response formula (such as Eq. (\ref{pumpingformula})) is derived by a field-theoretical approach,  evaluation of the pumped current can be carried out by calculating the retarded Green's function quantum mechanically.

The charge current is defined by 
\begin{align}
 j_i(\rv,t) &= -\frac{1}{2m}(\nabla_\rv-\nabla_{\rv'})_i\tr[ G^<(\rv,t,\rv',t)]|_{\rv'=\rv},
 \label{current}
\end{align}
where $G^<(\rv,t,\rv',t')\equiv i\average{c^\dagger_{\rv'}(t')c_\rv(t)}$ is the lesser Green's function, and $\tr$ is  the trace over the spin index. 
The spin current with spin polarization in the $\alpha(=x,y,z)$ direction is 
\begin{align}
 j_{{\rm s},i}^\alpha(\rv,t) &= -\frac{1}{4m}(\nabla_\rv-\nabla_{\rv'}) \tr[\sigma_\alpha G^<(\rv,t,\rv',t)]|_{\rv'=\rv}, \label{spincurrentdef}
\end{align}
where we include the magnitude of electron spin, $\frac{1}{2}$.
We assume that the Green's functions for the stationary potential $V_0$, denoted by $G_0$, are known.
The response to the dynamic potential $U$ is calculated by using Dyson's equation for the path-ordered Green's function defined for a complex time along a contour $C$ \cite{Haug07}
\begin{align}
 G(\rv,t,\rv',t')=G_0(\rv-\rv',t-t')+\int_c d t_1 \sum_{\rv_1} G_0(\rv-\rv_1,t-t_1)U(\rv_1,t_1)G(\rv_1,t_1,\rv',t').
\end{align}
The retarded component satisfies  
\begin{align}
 G^\ret(\rv,t,\rv',t')=G_0^\ret(\rv-\rv',t-t')+\int_{-\infty}^\infty d t_1 \sum_{\rv_1} G_0^\ret(\rv-\rv_1,t-t_1) U(\rv_1,t_1)G^\ret(\rv_1,t_1,\rv',t') ,
\end{align} 
which we write from now on by simply suppressing the space-time variables:
\begin{align}
 G^\ret=G_0^\ret+G_0^\ret U G^\ret =G_0^\ret+G^\ret U G_0^\ret . \label{Gret}
\end{align}
The advanced Green's function is the complex conjugate of the retarded Green's function in the Fourier representation.
The lesser component satisfies the  equation
\begin{align}
 G^<=G_0^<+G_0^\ret U G^< +G_0^< U G^\adv. \label{Gless}
\end{align}
The solution of Eq. (\ref{Gret}) is 
\begin{align}
G^\ret&= G_0^\ret\frac{1}{1-U G_0^\ret}= \frac{1}{1- G_0^\ret U}G_0^\ret=(1+G^\ret U) G_0^\ret, \label{Gzeroret} 
\end{align}
and then Eq. (\ref{Gless}) leads to 
\begin{align}
G^<&=(1+G^\ret U) G_0^<(1+ U G^\adv). \label{Glessfull}
\end{align}
We focus on the response linear in $U$, namely, use approximate expression of 
\begin{align}
G^<& \simeq G_0^<+ G_0^\ret U G_0^< +G_0^< U G^\adv. \label{Glesslienar}
\end{align}
For a static potential, we have 
\begin{align}
G_0^< &=  F(G_0^\adv - G_0^\ret), \label{Gzerolessres}
\end{align}
where  $F$ represents the Fermi distribution function (generally a matrix in spin space) in the scattering region.
The linear contribution of $G^<$, denoted by $\delta G^<$ is therefore
\begin{align}
\delta G^< 
 &= G_0^\ret (UF-FU) G_0^\adv + G_0^\adv FU G_0^\adv -G_0^\ret UF G_0^\ret . \label{delG}
\end{align}
We therefore  obtain a linear response formula for the charge current as 
\begin{align}
 j_i (\rv,t) &= -\frac{1}{2m}(\nabla_\rv-\nabla_{\rv'})_i\tr[\delta G^<(\rv,t,\rv',t)]|_{\rv'=\rv} \nnr
 &= -\frac{1}{2m}(\nabla_\rv-\nabla_{\rv'})_i
  \tr\lt[  G_0^\ret [U,F] G_0^\adv   
  + G_0^\adv F U G_0^\adv -G_0^\ret U F G_0^\ret  \rt] (\rv,t,\rv',t)|_{\rv'=\rv} .
 \label{spincurrent2}
\end{align}

\subsection{Pumping formula}

We estimate the asymptotic behavior of the pumped  current, i.e., the behavior in the region far away from the scattering region, at $|x|\ra \infty$.
The asymptotic behaviors of the Green's functions are 
\begin{align}
  G_0^\ret(\rv,\rv')|_{|x|\ra\infty} & \propto e^{ik|x|},
\end{align}
and $G_0^\adv(\rv,\rv')|_{|x|\ra\infty}  \propto e^{-ik|x|}$, where $k=\sqrt{2mE}$ is the asymptotic wave vector ($E$ is the energy of an asymptotic electron). 
The product of the Green's functions thus behaves asymptotically as (defining $W\equiv [U,F]$) 
\begin{align}
  \lim_{|x|\ra\infty}\int d\rv'  G_0^\ret(\rv,\rv') W(\rv') G_0^\adv (\rv',\rv) 
  & \propto e^{ik|x|} e^{-ik|x|} \nnr
  \lim_{|x|\ra\infty} \int d\rv' G_0^\ret(\rv,\rv') W(\rv') G_0^\ret (\rv',\rv) 
  & \propto e^{2ik|x|} ,
\end{align}
and therefore the contributions from the retarded (or advanced) Green's functions lead to only a rapidly oscillating asymptotic current and are neglected.
Note that in the linear response calculation assuming spatial uniformity but having finite-frequency external perturbation, such purely retarded or advanced contributions lead to finite (but usually not dominant) contributions.   
The asymptotic charge current obtained from Eq. (\ref{spincurrent2}) is therefore   
\begin{align}
 j_i (\rv,t)|_{x\ra\pm\infty} 
 &= \mp \frac{i k_i }{m}
\int dt' \int d\rv'   \tr \lt[ 
G_0^\ret(\rv,\rv',t,t') [U(\rv',t'),F] G_0^\adv (\rv',\rv,t',t)
  \rt],
 \label{pumpingformula}
\end{align}
with the spatial dependences explicitly recovered.
This pumping formula is one of the main results of this paper, and it represents  a linear response result of the external perturbation $U$ not restricted to slowly varying cases.

It is clear from the above formula that a current is generated  if the driving potential $U$ and the distribution function in the interaction region $F$ do not commute.
In other words, the particle distribution must change after the application of external driving potential for a finite current to arise.
The non commutativity vanishes in the adiabatic limit in the most strict sense, i.e., when the external potential is diagonal and its frequency $\omega$ is zero, resulting in a vanishing charge current proportional to $\Omega$ in the slowly varying limit \cite{Buttiker94} (see also Eq. (\ref{jpumpdynamic})). 
Instead, the pumped charge integrated over a period of external perturbation is finite (and often quantized \cite{Thouless83}) in the adiabatic pumping limit.
In the spin pumping case, by contrast, the non-commutativity necessary for pumping arises because of the 
spin-mixing nature of the dynamic magnetization, and a finite spin current proportional to the spin gauge field is pumped, as we shall see in Sec. \ref{SEC:sp}. 
The physical mechanisms of adiabatic charge pumping and spin pumping are therefore distinct when looked at from the viewpoint of the driving potentials.

\subsection{Relation to scattering approach}
The retarded and advanced Green's functions are proportional to the free functions multiplied by the transmission amplitude.
We see this following Ref. \cite{Fisher81} considering the one-dimensional case for simplicity.
The wave function of the system in the absence of the driving potential is
\begin{align}
 \Psi=\phi+G_0^\ret V_0 \phi,
\end{align}
where $\phi=e^{ikx}$ is an incoming wave.
The product $G^\ret V_0$ is written by use of $G_0^\ret V_0=G_0^\ret(g^\ret)^{-1}-1$ (derived from Eq. (\ref{Gzeroret})) as (recovering spatial coordinates)
\begin{align}
 \int dx' G_0^\ret(x,x') V_0(x')\phi(x') 
  &= \int dx' \lt[ -\delta(x-x')+G_0^\ret(x,x') \lt(\omega+\frac{\nablal_{x'}^2}{2m}\rt)\rt]\phi(x') ,
\end{align}
where $\omega=\frac{k^2}{2m}$.
Using integral by parts, we rewrite the  $\nablal^2$ term as
\begin{align}
  \int dx' G_0^\ret(x,x') \frac{\nablal_{x'}^2}{2m}\phi(x') 
     &= \frac{1}{2m}  [(\nabla_{x'}-ik)G_0^\ret(x,x')] \phi(x') |_{-\infty}^{x'=\infty}
      -\omega \int dx' G_0^\ret(x,x') \phi(x'),
\end{align}
resulting in a useful identity 
\begin{align}
 \int dx' G_0^\ret(x,x') V_0(x')\phi(x') 
  &= -\phi(x) +
  \frac{1}{2m}  [(\nabla_{x'}-ik)G_0^\ret(x,x')] \phi(x') |_{-\infty}^{x'=\infty}.\label{GVpres}
\end{align}
Since the retarded Green's function describes an out going wave,  $G_0^\ret(x,x') |_{x'\ra \pm\infty}\propto e^{\pm ikx'}$,  we see that only the contribution of $x'=-\infty$ survives (and not the one of $x'=\infty$) in Eq. (\ref{GVpres}).
The total wave function is thus 
\begin{align}
\Psi(x)
  &=   \frac{ik}{m} G_0^\ret(x,x') \phi(x') |_{x'=-\infty}. \label{PsiGresult}
\end{align}
The asymptotic behaviors of the wave function are  
written in terms of transmission amplitudes $t$  as 
$ \Psi(x\ra \infty)=t \phi(x)$, and we see therefore  that 
\begin{align}
 t &= \frac{ik}{m}G_0^\ret(x,x') \phi^*(x)\phi(x') |_{x=\infty,x'=-\infty}  .
\label{trresult}
\end{align}

\subsection{Case of a dynamic potential}
Here we present an example of the application of formula (\ref{pumpingformula}) in the case of the time-dependent potential discussed in Refs. \cite{Buttiker94,Moskalets02}.
For simplicity, we consider a one-dimensional spinless system. The potential $U$ is chosen as $U(x,t)=iu(x) e^{i\Omega t}$, where $u(x)$ is a localized function describing the potential profile and $\Omega $ is the external angular frequency.
The asymptotic pumped current at $|x|=\infty$ reduces to 
\begin{align}
 j (t) 
 &= \frac{k}{m}e^{i\Omega t}
  \sumom \int d\rv' \tr \lt[ 
G_0^\ret(\rv,\rv',\omega+\Omega)  [u(\rv')F(\omega) -F(\omega+\Omega)u(\rv')] G_0^\adv (\rv',\rv,\omega)
  \rt],
 \label{current4}
\end{align}
namely the current arises from the dynamic scattering potential modifies the electron's angular frequency (from $\omega$ to $\omega+\Omega$). 
In the frequency representation it is 
\begin{align}
 j (t)
 &= \frac{k}{m} 
  \sumom e^{i\Omega t} [f(\omega+\Omega)-f(\omega)] \int dx' u(x')  
G_0^\ret(x,x',\omega+\Omega)G_0^\adv   (x',x,\omega). \label{jpumpdynamic}
\end{align}
Using the relations (\ref{trresult}), the pumped current (\ref{jpumpdynamic}) is proportional to the transmission amplitude squared.
This expression corresponds to the one written in terms of scattering matrix and difference of the distribution function of the initial and  the excited states (like Eq. (8) of Ref. \cite{Moskalets02}).
While the expression in terms of scattering matrix is convenient to see the topological meaning \cite{Brouwer98}, the present linear response expression in terms of the Green's functions seems to be convenient for practical calculations. 
As is obvious from Eq. (\ref{jpumpdynamic}), a dynamic potential (finite $\Omega$) is necessary to generate a current in this spinless case with diagonal scattering potential.
In the slowly varying limit, the pumped current vanishes proportional to $\Omega$.

\section{Spin pumping by a dynamic magnetization}
\label{SEC:sp}
\subsection{Hamiltonian of minimum model}

In the previous section, we revisited a general theory of pumping. 
We now proceed to study a spin pumping effect driven by a dynamic magnetization. 
Usually, the  spin pumping effect is studied in the case of a ferromagnet with a spatially uniform magnetization. 
Thus, the effect is modeled by a simple model of free electrons in leads attached to a ferromagnetic dot, which we call the minimum model. 
Extension  to the case of a finite ferromagnet is straightforward.
\begin{figure}
	\includegraphics[width=0.4\textwidth]{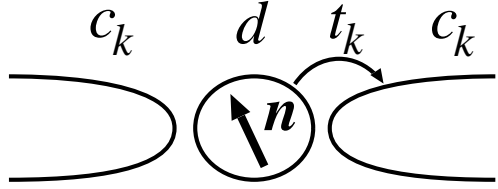}
	\caption{ Schematic  depicting the minimal model for spin pumping. 
	A ferromagnetic dot with magnetization direction $\nv$ is attached to leads. The electrons in the leads and the dot are denoted by $c_k$ and $d$, respectively, and the hopping amplitude between the dot and the lead is $t_k$. }
	\label{FIGdotmodel}
\end{figure}
The electron in the magnetic dot, represented by two-component operators $d(=(d_{\uparrow},d_{\downarrow})^{\rm t})$ ($^{\rm t}$ denotes transpose) and $d^\dagger $, is spin polarized because of the \sd exchange interaction with a time-dependent magnetization, whose direction is denoted by  a unit vector $\nv(t)$.
The magnetization is driven by an external magnetic field and is treated classically. 
The strength of the \sd exchange interaction is denoted by $M$. 
The total Hamiltonian is $H=H_{\rm d}+H_{\rm L}+H_{\hop}$, where 
\begin{align}
  H_{\rm d} &= \varepsilon_d d^\dagger d-M d^\dagger(\nv(t)\cdot\sigmav)d,
\end{align}
describes the electron in the dot, $\varepsilon_d$ is the energy of an unpolarized electron in the dot, and
\begin{align}
  H_{\rm L} &= \sum_{\kv} \epsilon_{\kv} c^\dagger_{\kv} c_{\kv},
\end{align}
describes the lead, where the electrons  are treated as free electrons with energy $\epsilon_{\kv}$.
The hopping between the magnetic dot and the leads is represented by a spin conserving term 
\begin{align}
  H_{\hop} &= \sum_{\kv} \hop_{\kv}( c^\dagger_{\kv}d+d^\dagger c_{\kv}),
\end{align}
where $t_{\kv}$ is the hopping amplitude.
(We may consider a multi-leads case by including indices specifying the leads in the field operators.)
The Lagrangian for the system is
\begin{align}
  L &= i\lt[ d^\dagger \partial_t d+ \sum_{\kv} c^\dagger_{\kv}\partial_t c_{\kv}\rt]-H. 
\end{align}

Throughout the paper we consider the case of strong \sd exchange interaction (large $M$) and diagonalize it by a unitary transformation of $d$ electron in the spin space \cite{TKS_PR08}.
The $d$ electron operator in the rotated frame is $\dtil\equiv U^{-1}(t)d$, where $U(t)$ is a $2\times2$ unitary matrix satisfying $U^{-1}(\nv\cdot\sigmav)U=\sigma_z$.
We can explicitly choose $U=\mv\cdot\sigmav$, where 
$\mv\equiv \lt(\sin\frac{\theta}{2}\cos\phi, \sin\frac{\theta}{2}\sin\phi, \cos\frac{\theta}{2}\rt)$, $\theta$ and $\phi$ are the polar coordinates of $\nv$.
The Lagrangian in the rotated frame reads
\begin{align}
  L &=  i\lt[ \dtil^\dagger \partial_t \dtil+ \sum_{\kv}c^\dagger_{\kv}\partial_t c_{\kv}\rt]-\tilde{H}, 
\end{align}
where
\begin{align}
  \tilde{H} &= \dtil^\dagger(\varepsilon_d-M\sigma_z) \dtil
   + \sum_{\kv} \epsilon_{\kv} c^\dagger_{\kv} c_{\kv}
   + \sum_{\kv} \hop_{\kv}( c^\dagger_{\kv}U\dtil+\dtil^\dagger U^{\dagger}c_{\kv})
   + \dtil^\dagger A_{{\rm s}}(t) \dtil,
\label{Hrotated}
\end{align}
where $ A_{{\rm s}}\equiv -iU^\dagger \partial_t U\equiv \sum_{\alpha} A_{{\rm s}}^\alpha \sigma_\alpha$ ($\alpha=x,y,z$) are the time components of a spin gauge field.  
(The spin gauge field has another space-time suffix representing the direction of flow \cite{TKS_PR08}, but here we suppress the suffix as we are interested only in the temporal direction.) 
These components are given more explicitly as  
\begin{align}
  \Av_{{\rm s}} =&
  \frac{1}{2} \lt( \begin{array}{c} -\dot{\theta}\sin\phi-\sin\theta\cos\phi\dot{\phi} \\ \dot{\theta}\cos\phi-\sin\theta\sin\phi\dot{\phi} \\ (1-\cos\theta)\dot{\phi} \end{array} \rt).
\end{align}
Equation (\ref{Hrotated}) indicates that the effect of dynamic magnetization is now represented by an effective magnetic field $A_{{\rm s}}^\alpha(t)$ along the direction $\alpha$.
The field $A_{{\rm s}}^\alpha(t)$ is a spin-dependent scalar potential for the electron, and it induces a spin-dependent shift of chemical potential; it is the so called  
\textquoteleft spin chemical potential\textquoteright.

It is clear from Eq. (\ref{Hrotated}) that the system is further simplified by applying the same unitary transformation to the electrons in the leads, i.e., by introducing $\ctil_{\kv}\equiv U^{\dagger} c_\kv$. The Hamiltonian then reads 
\begin{align}
  \tilde{H} &= \dtil^\dagger(\varepsilon_d-M\sigma_z) \dtil
   + \sum_{\kv} \epsilon_{\kv} \ctil^\dagger_{\kv} \ctil_{\kv}
   + \sum_{\kv} \hop_{\kv}( \ctil^\dagger_{\kv}\dtil+\dtil^\dagger \ctil_{\kv})
   + \dtil^\dagger A_{{\rm s}}(t) \dtil.
\label{Hrrotated}
\end{align}
In terms of $\dtil$ and $\ctil$ electrons, the system  reduces to an electron system having uniform spin polarization and spin-dependent scattering potential $A_{\rm s}$ inside the dot.
Because we are interested in the spin current at the linear order in the time-derivative of the magnetization, it is sufficient to treat the spin gauge field $A_{{\rm s}}$ as a time-independent potential (Note that the spin gauge field contains already a first-order derivative.) 
Therefore, the spin pumping system is equivalent to the one described by a Hamiltonian (\ref{Hgeneral}) with static but spin-dependent potentials,
\begin{align}
 V_0 &= v(\rv)[\delta \epsilon-M\sigma_z] \nnr
 U &= v(\rv) \sum_\alpha A_{\rm s}^\alpha \sigma_\alpha ,
\end{align}
where $v(\rv)$ is a function specifying the dot region 
($v(\rv)=1$ inside the dot and $v(\rv)=0$ outside) 
and $\delta \epsilon=\epsilon_d-\ef$ is the energy difference between the dot and the lead electron.

In the next subsection, we calculate the spin current by use of our pumping formula.
An approach evaluating the lesser Green's function without using the formula is presented in  Appendix \ref{APP:Keldysh} for comparison.

\subsection{Pumped spin current}

The spin pumping effect in the system described in Eq. (\ref{Hrrotated}) is now simply calculated  by applying our pumping formula (\ref{pumpingformula}).
The spin current we are interested in is obtained by inserting a Pauli matrix in the trace as (including magnitude of spin $\frac{1}{2}$)
\begin{align}
 \widetilde{j}_{{\rm s},i}^\alpha (\rv,t)|_{x\ra\pm\infty} 
 &= \mp \frac{i k_i }{2m}
  \int dt'\int d\rv' \tr \lt[ \sigma_\alpha 
G_0^\ret(\rv,\rv',t,t') [U(\rv',t'),F] G_0^\adv (\rv',\rv,t',t)
  \rt]|_{x\ra\pm\infty} .
 \label{spincurrent}
\end{align}
Note that the spin current here (denoted by $\widetilde{j}_{{\rm s}}$) is the one in a rotated frame (for $\ctil$ electrons).
An important observation in the present spin pumping case is that the spin current is generated even when the external perturbation $A_{\rm s}$ and  $U$ are static, since the potential and distribution function $F$ are spin-dependent matrices and thus $[U,F]$ is finite.
Defining $F=f_0+f_1\sigma_z$, with $f_0\equiv \frac{1}{2}\sum_{\pm}f_{\pm}$ and  
 $f_1 \equiv \frac{1}{2}\sum_{\pm}(\pm) f_{\pm}$, where $f_\pm$ is the Fermi distribution inside the dot for spin $\pm$.
The commutator in Eq. (\ref{spincurrent}) reads 
\begin{align}
 [U,F](\rv)
 &=-(f_{+}-f_{-}) v(\rv)\sum_{\pm} (\pm) A_{\rm s}^{\mp} \sigma_{\pm}
 =-2iv(\rv)f_1 \sigmav\cdot(\hat{\zv}\times \Av_{\rm s}) , \label{UFcomspin}
\end{align} 
where $A_{\rm s}^{\pm}\equiv A_{\rm s}^{x} \pm i A_{\rm s}^{y} $. 
Using
\begin{align}
 \tr[\sigma_\alpha G_0^\ret \sigma_\beta G_0^\adv] 
 &=\delta_{\alpha\beta}\sum_{\pm}G_{0,\pm}^\ret G_{0,\mp}^\adv -\epsilon_{\alpha\beta z} \sum_{\pm}(\pm i)G_{0,\pm}^\ret G_{0,\mp}^\adv,
\end{align}
for $\beta\neq z$, the spin current at $x=\infty$ is obtained as 
\begin{align}
 \widetilde{j}_{{\rm s},i}^\alpha (x) |_{x\ra\infty} 
 &= - \frac{k_i}{m}\sum_\beta A_{\rm s}^\beta\sumom  \int d\rv' (f_{+}(\omega)-f_{-}(\omega))  v(\rv') \nnr
 & \times 
 [-\mu_1(\rv,\rv',\omega) (\delta_{\alpha\beta}-\delta_{\alpha z}\delta_{\beta z})
  -\mu_2(\rv,\rv',\omega) \epsilon_{\alpha\beta z}],
\end{align} 
where
\begin{align}
\mu_1(\rv,\rv',\omega) & \equiv \frac{1}{2}\sum_{\pm}(\pm i)G_{0,\pm}^\ret(\rv,\rv',\omega) G_{0,\mp}^\adv(\rv',\rv,\omega) = -\Im [G_{0,+}^\ret(\rv,\rv',\omega) G_{0,-}^\adv(\rv',\rv,\omega) ] \nnr
\mu_2(\rv,\rv',\omega) & \equiv  \frac{1}{2}\sum_{\pm}G_{0,\pm}^\ret(\rv,\rv',\omega) G_{0,\mp}^\adv(\rv',\rv,\omega)
= \Re [G_{0,+}^\ret(\rv,\rv',\omega) G_{0,-}^\adv(\rv',\rv,\omega) ].
\label{musdef}
\end{align}
This is a spin current in the rotated frame. 
The spin current in the laboratory frame is 
\begin{align}
j_{{\rm s},i}^{\alpha}=\sum_\beta R_{\alpha\beta} \widetilde{j}_{{\rm s},i}^{\beta}.
\end{align}
Using
\begin{align}
 \sum_\beta R_{\alpha\beta}A_{{\rm s},0}^\beta =& -\frac{1}{2}(\nv\times\dot{\nv})_\alpha  + A_{{\rm s},0}^z\nv_\alpha \nnr
 \sum_{\beta \gamma} \epsilon_{\beta\gamma z}R_{\alpha\beta} A_{{\rm s},0}^\gamma 
  =& -\frac{1}{2} \dot{\nv}_\alpha,
\end{align}
 the asymptotic  pumped spin current in the laboratory frame is finally obtained as 
\begin{align}
 j_{{\rm s},i}^{\alpha} |_{x\ra \infty} 
 &= -\frac{k_i}{2m} \sumom \int d\rv'(f_{+}(\rv',\omega)-f_{-}(\rv',\omega))   v(\rv') 
[\mu_1(\nv\times\dot{\nv}) + \mu_2 \dot{\nv}]_\alpha. \label{jresultmu}
\end{align} 
(We introduced a position dependence in the Fermi distribution functions $f_{\pm}$ to be applicable to the case of a finite-size ferromagnet.) 
The pumped current at $x=\infty$ is thus written as 
\begin{align}
\jv_{{\rm s}}
 &= g_1(\nv\times\dot{\nv}) + g_2 \dot{\nv},\label{jsasymmptotic}
\end{align} 
where
\begin{align}
 g_1 &=\Re \eta, \;\;\;  g_2 =\Im\eta.
\end{align}
Here the spin pumping efficiency (corresponding to the spin mixing conductances of Ref. \cite{Tserkovnyak02})
is defined as 
\begin{align}
 \eta & \equiv  \frac{im}{2k}\sumom \int d\rv' (f_{+}(\omega)-f_{-}(\omega)) 
  \widetilde{t}_+(\omega,\rv') v(\rv')\widetilde{t}_-^*(\omega,\rv') , \label{spinpumpefficiency}
\end{align}
where 
\begin{align}
 \widetilde{t}_\pm(\omega,\rv')  & \equiv -\frac{ik}{m}G_{0,\pm}^\ret(\infty,\rv',\omega),
\end{align}
is an effective transmission amplitude connecting infinity and scattering region ($\rv'$).
The result (\ref{spinpumpefficiency}) is essentially the same as the one derived in Ref. \cite{Chen15}. 

In the present calculation, the mechanism of spin pumping is clearly identified to be the spin gauge field causing a spin mixing. 
In this sense the spin pumping effect is a non-adiabatic effect, if one defines the adiabaticity strictly to mean the case in which spin interaction is perfectly diagonal.

We note that, in contrast to the spin current, charge current is not pumped in the present situation of a static spin gauge field, as Eq. (\ref{UFcomspin}) leads to 
$\tr[ G_0^\ret(\rv,\rv_1,\omega) \sigma_{\pm} G_0^\adv   (\rv_1,\rv,\omega)]=0$.

An approach carrying out the calculation of the lesser Green's function without using a linear response formula is presented in Appendix \ref{APP:Keldysh}, in which the result for the asymptotic spin current  agrees with the above result.

\section{Long range spin pumping by diffusion}

The result (\ref{jsasymmptotic}) obtained in the previous section is for the asymptotic spin current, and is applicable to the case of a junction if the system is clean.
In reality, electron Green's functions have finite lifetimes because elastic scatterings result in a short-ranged propagation decaying in the length scale of mean free path, $\ell$.
The magnitude of the pumped spin current near the interface is governed by  spatial averages of the  coefficients $\mu_i$ ($i=1,2$), 
$\bar\mu_i\equiv  \int d\rv\mu_i(\rv,\rv')$. 
(The spatial integral can be extended to  infinity, as only contributions from near the interface dominate, owing to  the short-ranged nature of  the Green's functions.)
If the Green's function near the interface is approximated by a free function with a  constant spin polarization $\bar{M}$, i.e., 
$G_{0,\pm}^\ret(\rv,\rv',\omega)=\frac{1}{V}\sumkv \frac{e^{i\kv(\rv-\rv')}}{\omega -\ekv\pm\bar{M}+i\eta}$, 
we have (for small $\omega$ and for $\rv'$ in the scattering region)
\begin{align}
\int d\rv G_{0,\pm}^\ret(\rv,\rv',\omega)G_{0,\mp}^\adv(\rv',\rv,\omega)
&=
\frac{1}{V}\sumkv G_{0,\pm}^\ret(\kv,\omega)G_{0,\mp}^\adv(\kv,\omega)
\nnr
& \simeq \pm i\pi\frac{\bar{\dos}}{\bar{M}}\frac{1}{1\pm i \eta/\bar{M}}
\;\;\;(\bar{M}\tau\gg1),
\label{GrGainterface}
\end{align}
where $\bar{\dos}\equiv\frac{1}{2}(\dos_++\dos_-)$, $\dos_\pm$ is the density of states of electrons with spin $\pm$ near the interface. 
We therefore have from Eq.(\ref{musdef}) 
$\bar\mu_1=-\frac{\pi}{2}\frac{\bar{\dos}}{\bar{M}}$ and 
$\bar\mu_2= \frac{\pi}{2}\frac{\bar{\dos}\eta}{\bar{M}^2}$, 
where we ignored higher orders of $\eta/\bar{M}$.
Although the contribution  (\ref{jsasymmptotic}) is localized near the interface within the length scale of the elastic mean free path of the electron, there arises another diffusive contribution that survives for a longer length scale than the electron's mean free path.

Below, we  study the injection of a diffusive spin current injected at a junction of a ferromagnetic metal and a dirty nonmagnetic metal.  
(A Green's function approach to the diffusive spin current was briefly discussed  in Ref. \cite{Chen15}.) 
The diffusive contribution arises by including the elastic scattering by random impurities, which we assume to exist in a nonmagnetic metal \cite{Abrikosov75}.
The impurity is modeled by a $\delta$-function potential,
\begin{eqnarray}
 V_{\rm i}(\rv) &= \sum_{i}^{N_{\rm i}}v_{\rm i}\delta(\rv-\Rv_{i}),
\end{eqnarray}
where $N_{\rm i}$ and $\Rv_{i}$ are the total  number of impurities, with the positions of the impurities labeled by  the respective $i$, and  $v_{\rm i}$ represents the strength of the potential. 
The average of the impurity positions is determined as 
$\sum_{ij}\average{e^{i\qv\cdot\Rv_i}e^{i\qv'\cdot\Rv_j} }=N_{\rm i}\delta(\qv+\qv')$.
The impurity scattering leads to a finite lifetime for the electron Green's function, 
$\tau_0\equiv (2\pi\dos n_{\rm i}v_{\rm i}^2)^{-1}$, where $n_{\rm i}\equiv N_{\rm i}/N$ ($N$ is the total number of  lattice sites) is the impurity density, and we neglect the spin dependence of the density of states.
The impurity scattering also gives rise to the so called vertex corrections, which are given by  multiple electrons'  retarded and advanced Green's functions  representing  multiple scatterings.
The asymptotic spin current in the rotated frame including $n$ scatterings is
\begin{align}
 \widetilde{j}_{{\rm s},i}^{(n),\alpha} (\rv)
 &= -\frac{i}{4m}(\nabla_{\rv}-\nabla_{\rv''})_i \sumom 
\prod_{j=1}^{n}\lt[\int d\rv_j\rt] (n_{\rm i}v_{\rm i}^2)^n W_\mp(\rv') \nnr 
 & \times \tr \lt[ \sigma_\alpha 
\prod_{j=0}^{n}[ G_0^\ret(\rv_j,\rv_{j+1},\omega)]\sigma_\pm 
\prod_{j'=0}^{n} [G_0^\adv (\rv_{j'+1},\rv_{j'},\omega)]
  \rt],
 \label{spincurrent_ndiffusion}
\end{align}
where $\rv_0\equiv \rv$, $\rv_{n+1}\equiv \rv'$, $\nabla_{\rv''}$ applies to $\rv$ in the advanced Green's function,  and $W_\pm\equiv \tr[\sigma_\pm[U,F]]=\pm(f_+-f_-)v A_{\rm s}^\pm$.
The expression (\ref{spincurrent_ndiffusion}) describes the diffusion process of the spin polarization induced by the interaction $W_\mp(\rv') $, and it shows that multiple scattering is necessary as  
each scattering leads to the contribution of the order of unity when the spin splitting is neglected.
In fact, a spatial integral of a pair of Green's functions with small wave vector $\qv$ gives
\begin{align}
 n_{\rm i}v_{\rm i}^2 \int d\rv e^{i\qv\cdot\rv} G_{0,\pm}^\ret(\rv,\rv',\omega)G_{0,\mp}^\adv(\rv',\rv,\omega) 
 &=\frac{1}{V}\sumkv e^{i\qv\cdot\rv'}G_{0,\pm}^\ret(\kv,\omega)G_{0,\mp}^\adv(\kv+\qv,\omega) \nnr 
 & \simeq 1-Dq^2\tau  \;\;\;(\bar{M}\tau\ll1),
 \label{tothediffusionpole}
\end{align} 
where $D$ is the diffusion constant if the spin splitting is neglected, i.e., in a nonmagnetic dirty metal. The summation over $n$ therefore leads to a well-known diffusion pole:
$\sum_{n=0}^\infty (1-Dq^2\tau )^n=\frac{1}{Dq^2\tau}$.
By contrast, the Green's functions near the interface where spin splitting becomes essential lead to a small contribution of $ n_{\rm i}v_{\rm i}^2 \int d\rv G_{0,\pm}^\ret(\rv,\rv',\omega)G_{0,\mp}^\adv(\rv',\rv,\omega)
 =O( \frac{1}{\bar{M}\tau_0})$, 
as seen from Eq. (\ref{GrGainterface}).
This contribution is therefore taken into account to the linear order.
The diffusive spin current obtained as a sum of the $n$-th order contributions is therefore  
\begin{align}
 \widetilde{j}_{{\rm s},i}^{{\rm (D)},\pm} (\qv) 
 &\equiv \sum_{n=0}^\infty 
 \widetilde{j}_{{\rm s},i}^{(n),\alpha} 
 = -i \frac{q_i}{4m}\frac{n_{\rm i}v_{\rm i}^2}{Dq^2\tau}\sumom \int d\rv'\int d\rv'' W_\pm(\rv') G_0^\ret(\rv'',\rv',\omega) G_0^\adv (\rv',\rv'',\omega),
 \label{spincurrent_ndiffusion2}
\end{align}
where the superscript $\pm$ represents the transverse components and the $z$ component vanishes as is the case for the local contribution  (Eq.  (\ref{jsasymmptotic})).
The integration over $\rv'$ is in the scattering region, where $W_\pm(\rv')$ is finite.
In the real space representation, it is 
\begin{align}
 \widetilde{j}_{{\rm s},i}^{{\rm (D)},\alpha} (\rv) 
 &= -D\nabla_i \int d\rv'' D_0(\rv-\rv'') \frac{n_{\rm i}v_{\rm i}^2}{4mD}\sumom \int d\rv' (f_{+}(\omega)-f_{-}(\omega))   v(\rv') \nnr
 & \times \lt[ A_{\rm s}^\alpha \Re[ i G_0^\ret(\rv'',\rv',\omega) G_0^\adv (\rv',\rv'',\omega)]
       + (\hat{\zv}\times\Av_{\rm s})^\alpha \Im[ i G_0^\ret(\rv'',\rv',\omega) G_0^\adv (\rv',\rv'',\omega)] \rt]
  ,
 \label{spincurrent_diffusion1}
\end{align}
where 
\begin{align}
  D_0(\rv-\rv') \equiv \sumqv \frac{e^{i\qv\cdot(\rv-\rv')}}{Dq^2\tau},
\end{align}
is the diffusion propagator.
When we take account of spin relaxation owing to magnetic impurities or spin-orbit interaction,
the diffusion propagator is modified to be the massive propagator 
\begin{align}
  D(\rv-\rv') \equiv \sumqv \frac{e^{i\qv\cdot(\rv-\rv')}}{Dq^2\tau+\eta_\perp},
  \label{Diffusiondef}
\end{align}
where $\eta_\perp$ denotes the dimensionless spin relaxation rate.
The diffusive spin current in the laboratory frame is therefore obtained as 
\begin{align}
 {\jv}_{{\rm s},i}^{{\rm (D)}}
 &= -D\nabla_i \rhov_{\rm s}, \label{diffusivespincurrent}
\end{align}
 where 
\begin{align}
 \rho_{\rm s}(\rv) &\equiv 
  \int d\rv'' D(\rv-\rv'') 
   \biggl[ \Re[\bar{\eta}(\rv'')] (\nv\times\dot{\nv})+  \Im[\bar{\eta}(\rv'')]\dot{\nv} \biggr ],  
   \label{diffusivespindensity}
\end{align}
   is the diffusively induced non-equilibrium spin accumulation and 
\begin{align} 
\bar{\eta}(\rv'') &\equiv 
 \frac{n_{\rm i}v_{\rm i}^2}{4mD} \sumom \int d\rv' (f_{+}(\omega)-f_{-}(\omega))   v(\rv') 
 i G_0^\ret(\rv'',\rv',\omega) G_0^\adv (\rv',\rv'',\omega)
, \label{spincurrent_diffusion2}
\end{align}
is a parameter representing the spin pumping efficiency near the interface.
The result of Eq. (\ref{diffusivespincurrent}) indicates that the long-ranged component of spin current is determined by the spin accumulation profile, supporting the scenario of Silsbee \cite{Silsbee79}. 
The magnitudes of the spin current and spin density are related by  
${j}_{{\rm s}}^{{\rm (D)}}=\rho_{\rm s} \frac{D}{\els}$.

The diffusion propagator (\ref{Diffusiondef}) satisfies the diffusion equation 
$(D\tau\nabla^2-\eta_\perp)D(\rv)=0$.
If we consider a nonmagnetic metal with length (perpendicular to the interface) $L$, a boundary condition of $\nabla_x D(x=L)=0$ is required as a result of vanishing spin current  at  the edge $x=L$, choosing $x$ axis perpendicular to the interface, resulting in a commonly used expression \cite{Tserkovnyak02b} of 
\begin{align}
D(x) &=D_0 \frac{\cosh((x-L)/\els)}{\sinh (L/\els)},
\end{align}
where $D_0$ is a coefficient and $\els\equiv \sqrt{D\tau\eta_\perp}$ is the spin diffusion length.

\section{Summary}
We have derived a general pumping formula to describe currents induced by a pumping potential.
The source for the pumping is represented by the non-commutativity between the pumping potential and particle distribution matrix.
The formula was applied to study the spin pumping by a uniform and dynamic magnetization.
Using a unitary transformation to diagonalize the \sd exchange interaction, the low-energy behavior of the magnetization was described in terms of the spin gauge field.
The spin pumping effect in the slowly varying limit was shown to be equivalent  in the rotated frame to a pumping effect caused by a static and spin-mixing  chemical potential, consisting of the two non-adiabatic components of the spin gauge field.
The results are consistent with a previous study assuming small-amplitude precession \cite{Chen15} and  conventional spin pumping theory \cite{Tserkovnyak02}.

As is seen in Eqs. (\ref{spincurrent_ndiffusion2})(\ref{spincurrent}), the dominant spin current contributions are products of retarded and advanced Green's functions. This fact is expected as such terms describe the non-equilibrium processes arising from excitations, as is known in linear response theory.
For such a non-equilibrium contribution to arise, non-commutativity of the pumping source and particle distribution ($[U,F]$ in Eq. (\ref{spincurrent})) is essential. 
If such non-commutativity is absent, the retarded and advanced components do not mix in the expression for the current; namely, the current is an equilibrium current.
The non-equilibrium nature of the pumping effect is essential also for the  diffusive contribution of the pumping.
This fact is reflected in the calculations by the appearance of a diffusion pole only when the retarded and advanced Green's functions are connected (as in Eq. (\ref{tothediffusionpole})).


\acknowledgements
GT thanks C. Uchiyama, K. Hashimoto, 
and H. Kohno for valuable discussions.
This work was supported by 
a Grant-in-Aid for Exploratory Research (No.16K13853) from the Japan Society for the Promotion of Science 
and  
a Grant-in-Aid for Scientific Research on Innovative Areas (No.26103006) from The Ministry of Education, Culture, Sports, Science and Technology (MEXT), Japan.

\appendix
\section{Approach by a direct calculation of lesser Green's function}
\label{APP:Keldysh} 
In the main text, we  carried out a calculation based on a quantum-mechanical picture of the scattering potential. 
Here, we present a formulation of a fully field-theoretical picture for comparison.
The current is calculated from the hopping amplitude between the dot and lead, and the hopping is treated to the second order.

The expressions for the charge and spin currents are derived by calculating the time derivative of the charge and spin densities, respectively.
The density operator of the dot is defined as 
$
\rho_\mu \equiv   d^\dagger \sigma_\mu d
$, 
where $\mu=0,x,y,z$ corresponds to charge ($\mu=0$) and three components of spin  ($\mu=x,y,z$)  and $\sigma_0\equiv1$.
Its derivative  is calculated from the commutator with the total Hamiltonian as 
\begin{align}
\frac{d \hat{\rho}_\mu}{dt} = & i  \lt( d^\dagger \sigma_\mu [H,d] + [H,d^\dagger] \sigma_\mu d \rt)  
\nnr
& \equiv \widehat{j}_{\mu},
\end{align}
where we define the current operator as the net current flowing into the dot.
The operators representing the charge and spin currents between the dot and the lead are
\begin{align}
  \widehat{j}^{(0)}_{\mu} &= \sum_{\kv} i\hop_{\kv}(c^\dagger_{\kv}\sigma_{\mu} d -d^\dagger \sigma_{\mu} c_{\kv}).
\label{jzerodef}
\end{align}

\subsection{Expression for the  current in the rotated frame}
We calculate the currents in the rotated frame based on the Hamiltonian (\ref{Hrotated}). 
The expectation values of the currents in the laboratory frame are
\begin{align}
 {\jzero_{\mu}} (t) & \equiv \sum_{\kv} \hop_{\kv} \tr \biggl[
  \sigma_\mu U(t) G_{dc_{\kv}}(t,t) - U^\dagger(t) \sigma_\mu G_{c_{\kv}d}(t,t)  \biggr]^<,
  \label{currentdef}
\end{align}
where $G_{dc_{\kv}}(t,t')^<\equiv i\average{ c_{\kv}^\dagger(t')\dtil (t) } $ and 
$G_{c_{\kv}\dtil}(t,t')^<\equiv i\average{ \dtil^\dagger(t')c_{\kv}(t) } $ are lesser components of path-ordered (Keldysh) Green's functions in the rotated $d$ electron.
Those Green's functions connecting the electrons in the dot and the lead satisfy Dyson's equation 
\begin{align}
  G_{dc_{\kv}}(t,t')
&=
  \hop_{\kv} \int_C dt_1 G_d(t,t_1) U^\dagger(t_1) g_{\kv}(t_1,t')
\nnr
  G_{c_{\kv}d}(t,t')
&=
  \hop_{\kv} \int_C dt_1 g_{\kv}(t,t_1) U(t_1) G_d(t_1,t') 
\end{align}
where $G_{d}(t,t')\equiv -i\average{ \dtil^\dagger(t')\dtil (t) } $ is the dot Green's function including the hopping and interactions, and $g_{\kv}(t,t')\equiv - i\average{ c_{\kv}^\dagger(t') c_{\kv} (t) }_0 $ denotes the free Green's function of the lead ($\average{\ }$ denotes the expectation value without hopping and interactions).
Their lesser components are 
\begin{align}
  G_{dc_{\kv}}^<(t,t')
& =
  \hop_{\kv} \int_{-\infty}^\infty dt_1 [G_d^\ret(t,t_1) U^\dagger(t_1) g_{\kv}^<(t_1,t') + G_d^<(t,t_1) U^\dagger(t_1) g_{\kv}^\adv(t_1,t')]
\nnr
  G_{c_{\kv}d}^<(t,t')
&=
  \hop_{\kv} \int_{-\infty}^\infty dt_1 [g_{\kv}^\ret(t,t_1)  U(t_1) G_d^< (t_1,t') + g_{\kv}^<(t,t_1) U(t_1) G_d^\adv (t_1,t') ],
\end{align}
where the lesser free Green's function of the lead is 
\begin{align}
  g^<_{\kv}(t,t') &= f_{\kv}[ g^\adv_{\kv}(t,t')-g^\ret_{\kv}(t,t') ] .
\end{align}
Here $ f_{\kv}$, $g^\ret_{\kv}$ and $g^\adv_{\kv}$ are the Fermi distribution function, and the retarded and advanced Green's functions of the lead, respectively.
The current is therefore (using the fact that lead Green's functions are not spin-polarized) 
\begin{align}
 {j}^{(0)}_{\mu} (t) & = \sum_{\kv} (\hop_{\kv})^2 \int_{-\infty}^\infty dt_1 
  \tr \biggl[ U^\dagger(t_1)\sigma_\mu U(t) 
  \biggl(G_d^\ret(t,t_1) g_{\kv}^<(t_1,t)+G_d^<(t,t_1) g_{\kv}^\adv(t_1,t) \biggr)\nnr 
&  - U^\dagger(t)\sigma_\mu U(t_1) \biggl( g_{\kv}^\ret(t,t_1) G_d^< (t_1,t) + g_{\kv}^<(t,t_1) G_d^\adv (t_1,t) ) 
  \biggr) \biggr].
  \label{current2}
\end{align}
As we are focusing on the adiabatic limit, we can carry out expansion 
\begin{align}
  U^\dagger(t_1)\sigma_\mu U(t) =& U^\dagger(t_1) U(t) U^\dagger(t) \sigma_\mu U(t)  
   = \biggl(1-i(t_1-t)A_{{\rm s},0}(t)+\cdots \biggr) R_{\mu\nu}(t)\sigma_\nu \nnr 
  U^\dagger(t)\sigma_\mu U(t_1) =& U^\dagger(t) \sigma_\mu U(t) U^\dagger(t) U(t_1)   
   = R_{\mu\nu}(t)\sigma_\nu \biggl(1+i(t_1-t)A_{{\rm s},0}(t)+\cdots \biggr) 
\end{align}
to obtain 
\begin{align}
 {j}^{(0)}_{\mu} (t) & =  R_{\mu\nu}(t)\sigma_\nu [  \tilde{\jzero}_{\nu}+ \delta{\jzero}_{\nu} ] \\
 \tilde{j}^{(0)}_{\mu} (t) & \equiv 
\sum_{\kv} (\hop_{\kv})^2 \int_{-\infty}^\infty dt_1 
  \tr \biggl[ \sigma_\mu 
  \biggl(G_d^\ret(t,t_1) g_{\kv}^<(t_1,t)+G_d^<(t,t_1) g_{\kv}^\adv(t_1,t) \nnr 
&  - \biggl( g_{\kv}^\ret(t,t_1) G_d^< (t_1,t) + g_{\kv}^<(t,t_1) G_d^\adv (t_1,t) ) 
  \biggr)\biggr) \biggr] \label{currentpara} \\
\delta{j}^{(0)}_{\mu} (t) & \equiv 
- \sum_{\kv} (\hop_{\kv})^2 A_{{\rm s},0}^\alpha (t)\int_{-\infty}^\infty dt_1 i(t_1-t) 
  \tr \biggl[ \sigma_\alpha \sigma_\mu 
  \biggl(G_d^\ret(t,t_1) g_{\kv}^<(t_1,t)+G_d^<(t,t_1) g_{\kv}^\adv(t_1,t) \biggr)\nnr 
&  + \sigma_\mu \sigma_\alpha \biggl( g_{\kv}^\ret(t,t_1) G_d^< (t_1,t) + g_{\kv}^<(t,t_1) G_d^\adv (t_1,t) ) 
  \biggr) \biggr]  
  \label{currentdia}
\end{align}
The first contribution, the 'paramagnetic' spin current in the rotated frame, is calculated to the linear order of the spin gauge field while the second contribution, the 'diamagnetic' current, is of linear order by definition.  

Taking account of a spin gauge field having in general a finite angular frequency, the Fourier transform of 
$G_{d}^{\ret}$ is 
$G_{d}^{\ret}(\omega,\omega+\Omega)=\int dt\int dt' e^{i\omega(t-t')}e^{-i\Omega t'} G_{d}^{(0),\ret}(t,t')$.
The current then reads 
\begin{align}
 \tilde{j}^{(0)}_{\mu} (t) & = \sum_{\kv} (\hop_{\kv})^2 \sumom\sumOm e^{i\Omega t}
  \tr \biggl[\sigma_\mu \biggl( G_d^\ret(\omega,\omega+\Omega) g_{\kv}^<(\omega+\Omega)
       +G_d^<(\omega,\omega+\Omega) g_{\kv}^\adv(\omega+\Omega)
\nnr 
&   - g_{\kv}^\ret(\omega) G_d^< (\omega,\omega+\Omega) - g_{\kv}^<(\omega) G_d^\adv (\omega,\omega+\Omega) 
  \biggr) \biggr].
  \label{currentFT}
\end{align}

\subsection{Paramagnetic contribution}
The Dyson's equation for the dot Green's function is  
\begin{align}
  G_d(t,t') &= G_{d}^{(0)}(t,t') + \int_C dt_2 G_{d}^{(0)}(t,t_2) 
     (\Av_{{\rm s},0}(t_2)\cdot\sigmav) G_{d}(t_2,t') ,
\end{align}
where $G_{d}^{(0)}$ is the dot Green's function including the hopping but without the spin gauge field.
Below we neglect contributions of the second and higher orders in the spin gauge field.
In the frequency representation, the retarded component is therefore
\begin{align}
  G_d^\ret(\omega,\omega+\Omega) 
  &= G_{d}^{(0),\ret}(\omega)\delta_{\omega,\omega'} + 
     G_{d}^{(0),\ret}(\omega) \sumOm(\Av_{{\rm s},0}(\Omega)\cdot\sigmav) G_{d}^{(0),\ret}(\omega+\Omega,\omega') ,\label{GdDysonomega}
\end{align}
where $ \Av_{{\rm s},0}(\Omega)\equiv  \int_{-\infty}^\infty dt \Av_{{\rm s},0}(t)e^{-i\Omega t}$.
The lesser component is similarly
\begin{align}
  G_d^< & (\omega,\omega+\Omega) 
  = G_{d}^{(0),<}(\omega)\delta_{\omega,\omega'} \nnr
  & + 
     G_{d}^{(0),\ret}(\omega) (\Av_{{\rm s},0}(\Omega)\cdot\sigmav) G_{d}^{(0),<}(\omega+\Omega,\omega') 
     +  G_{d}^{(0),<}(\omega) (\Av_{{\rm s},0}(\Omega)\cdot\sigmav) G_{d}^{(0),\adv}(\omega+\Omega,\omega') .\label{GdDysonomegaless}
\end{align}
In the adiabatic limit, the external angular frequency $\Omega$ is neglected in the Green's functions, resulting in a diagonal Green's function,i.e., 
$G_{d}^\ret(\omega,\omega')=G_{d}^\ret(\omega)\delta_{\omega,\omega'}$ 
(corresponding to the instantaneous approximation \cite{Zhou99}).
Therefore 
\begin{align}
  G_d^\ret(\omega,\omega+\Omega) 
  &\simeq 
  G_{d}^{(0),\ret}(\omega) + 
     G_{d}^{(0),\ret}(\omega) (\Av_{{\rm s},0}(\Omega)\cdot\sigmav) G_{d}^{(0),\ret}(\omega) \nnr
 G_d^<(\omega,\omega+\Omega) 
  &\simeq  G_{d}^{(0),<}(\omega) + 
     G_{d}^{(0),\ret}(\omega) (\Av_{{\rm s},0}(\Omega)\cdot\sigmav) G_{d}^{(0),<}(\omega) 
     +  G_{d}^{(0),<}(\omega) (\Av_{{\rm s},0}(\Omega)\cdot\sigmav) G_{d}^{(0),\adv}(\omega) .\label{GdDysonadiabatic}
\end{align}
The Green's function $G_{d}^{(0),<}(\omega)$ satisfies
\begin{align}
  G_{d}^{(0),<}(\omega) =F_d(\omega)[G_{d}^{(0),\adv}(\omega)-G_{d}^{(0),\ret}(\omega)],
\end{align}
where 
\begin{align}
 F_d(\omega)\equiv \lt(
 \begin{array}{cc} f_{d,+}(\omega) & 0 \\ 0 & f_{d,-}(\omega) \end{array} 
         \rt)       
\end{align}
is  a matrix of the Fermi distribution function of the dot.
Then 
\begin{align}
 G_d^<& (\omega,\omega+\Omega) 
  \simeq 
  F_d [G_{d}^{(0),\adv}(\omega) - G_{d}^{(0),\ret}(\omega) ]\nnr
&  + 
     G_{d}^{(0),\ret}(\omega) (\Av_{{\rm s},0}(\Omega)\cdot\sigmav) 
      F_d [G_{d}^{(0),\adv}(\omega) - G_{d}^{(0),\ret}(\omega) ]
     +   F_d [G_{d}^{(0),\adv}(\omega) - G_{d}^{(0),\ret}(\omega) ] (\Av_{{\rm s},0}(\Omega)\cdot\sigmav) G_{d}^{(0),\adv}(\omega) 
     \nnr
   = &  F_d [G_{d}^{(0),\adv}(\omega) - G_{d}^{(0),\ret}(\omega) ] \nnr
&  + A_{{\rm s},0}^\beta(\Omega)\biggl[
     G_{d}^{(0),\ret}(\omega)
     [\sigma_\beta, F_d] G_{d}^{(0),\adv}(\omega) 
     +   F_d G_{d}^{(0),\adv}(\omega) \sigma_\beta G_{d}^{(0),\adv}(\omega) 
     - G_{d}^{(0),\ret}(\omega) \sigma_\beta F_d G_{d}^{(0),\ret}(\omega) \biggr]
\end{align}

The paramagnetic current linear in the spin gauge field is  therefore 
\begin{align}
 \widetilde{\jzero_{\mu}} (t) & = \sum_{\kv} (\hop_{\kv})^2 \sumom A^\beta_{{\rm s},0}(t)
  \rho_\kv(\omega) \nnr
  \times i \tr & \biggl[\sigma_\mu
  \biggl[ 
  - G_d^{(0),\ret}(\omega)\sigma_\beta G_d^{(0),\ret}(\omega) (F_d-f_{\kv})
 +(F_d-f_\kv) G_d^{(0),\adv} \sigma_\beta G_d^{(0),\adv} (\omega) 
  + G_{d}^{(0),\ret}(\omega)
     [ \sigma_\beta, F_d] G_{d}^{(0),\adv}(\omega)  
  \biggr]\biggr],
  \label{currentFT3}
\end{align}
where $\rho_\kv(\omega)\equiv -i[g_{\kv}^\adv(\omega)-g_{\kv}^\ret(\omega)]=2\pi \delta(\omega-\ekv)$ 
( $g_{\kv}^<(\omega)=i f_\kv\rho_\kv(\omega)$).
Teh trace is evaluated using an identity 
\begin{align}
  \tr[\sigma_\alpha A\sigma_\beta B] 
  &= (\delta_{\alpha\beta}-\delta_{\alpha z}\delta_{\beta z})\sum_{\sigma=\pm} A_\sigma B_{-\sigma}
  + \delta_{\alpha z}\delta_{\beta z}\sum_{\sigma} A_\sigma B_{\sigma}
  -i\epsilon_{\alpha \beta z} \sum_{\sigma} \sigma A_\sigma B_{-\sigma},
\end{align}
for diagonal matrices $\displaystyle A\equiv \lt(\begin{array}{cc} A_+ & 0 \\ 0 & A_- \end{array}\rt)$ and $\displaystyle B\equiv \lt(\begin{array}{cc} B_+ & 0 \\ 0 & B_- \end{array}\rt)$.

The result of the 'paramagnetic' part of the current (Eq. (\ref{currentpara}) is  
\begin{align}
 \widetilde{\jzero_{\alpha}} (t) & = \sum_{\kv} (\hop_{\kv})^2 \sumom A^\beta_{{\rm s},0}(t)
  \rho_\kv(\omega) 
\biggl[ (\delta_{\alpha\beta}-\delta_{\alpha z}\delta_{\beta z})\tilde{\mu}_1^{\rm p}
 + \epsilon_{\alpha \beta z}\tilde\mu_2^{\rm p} 
 +  \delta_{\alpha z}\delta_{\beta z} \tilde\mu_3^{\rm p}
 \biggr],   \label{jtilderesultpara}
\end{align}
where (the superscript p denotes paramagnetic) 
\begin{align}
\tilde{\mu}_1^{\rm p}
& \equiv i \sum_{\sigma} (f_{d,\sigma}-f_{\kv})
 [G_{d,+}^{(0),\adv}(\omega) G_{d,-}^{(0),\adv}(\omega) - G_{d,+}^{(0),\ret}(\omega) G_{d,-}^{(0),\ret}(\omega) ] 
 \nnr
 &
 -i(f_{d+}-f_{d-}) [G_{d,+}^{(0),\ret}(\omega) G_{d,-}^{(0),\adv}(\omega) - G_{d,-}^{(0),\ret}(\omega) G_{d,+}^{(0),\adv}(\omega) ]  \nnr 
\tilde{\mu}_2^{\rm p}
& \equiv 
 (f_{d,+}-f_{d,-}) \lt[ ( G_{d,+}^{(0),\adv}(\omega) + G_{d,+}^{(0),\ret}(\omega))
(G_{d,-}^{(0),\adv}(\omega) +G_{d,-}^{(0),\ret}(\omega) )\rt]
\nnr
\tilde{\mu}_3^{\rm p}
& \equiv  i \sum_{\sigma} (f_{d,\sigma}-f_{\kv}) [(G_{d,\sigma}^{(0),\adv}(\omega))^2 - (G_{d,\sigma}^{(0),\ret}(\omega) )^2 ] .
\label{tildemuthree}
\end{align}

\subsection{Total pumped spin current}
The 'diamagnetic' current, (Eq. (\ref{currentdia})), is calculated using 
\begin{align}
\int_{-\infty}^\infty dt_1 i(t-t_1) e^{-i(\omega-\omega')(t-t_1)} 
= - \frac{d}{d\omega}\delta(\omega-\omega') 
\end{align}
as 
\begin{align}
\delta{j}^{(0)}_{\alpha} (t) & = 
 \sum_{\kv} (\hop_{\kv})^2 A_{{\rm s},0}^\beta (t)\sumom \rho_k(\omega) 
 [ (\delta_{\alpha\beta}-\delta_{\alpha z}\delta_{\beta z})\tilde\mu_1^{\rm d}
  -\delta_{\alpha z} \delta_{\beta z} \tilde{\mu}_3^{\rm p}],
\end{align}
where
\begin{align}
\tilde\mu_1^{\rm d} \equiv & i\sum_{\sigma}(f_k-f_{{\rm d},\sigma})((G_{d,\sigma}^\adv(\omega))^2 - (G_{d,\sigma}^\ret(\omega))^2 ) ,
\end{align}
where we have neglected the higher orders in $\hbar/(\ef\tau)$.
The total spin current in the rotated frame, the sum of the paramagnetic and diamagnetic contributions, is therefore
\begin{align}
 \widetilde{j}^{(0)}_{\alpha}+\delta{j}^{(0)}_{\alpha} 
 & = \sum_{\kv} (\hop_{\kv})^2 \sumom A^\beta_{{\rm s},0}(t)
  \rho_\kv(\omega) 
\biggl[ (\delta_{\alpha\beta}-\delta_{\alpha z}\delta_{\beta z}){\tilde{\mu}}_1
 + \epsilon_{\alpha \beta z}\tilde\mu_2
 \biggr],   \label{jtilderesult}
\end{align}
where  $\tilde{\mu}_2 \equiv \tilde{\mu}_2^{\rm p}$ and 
\begin{align}
\tilde{\mu}_1 \equiv \tilde\mu_1^{\rm p} + \tilde\mu_1^{\rm d} 
&=
i \sum_{\sigma} \biggl[
-(f_{d,\sigma}-f_{\kv})
 [G_{d,\sigma}^{(0),\adv}(\omega) (G_{d,\sigma}^{(0),\adv}(\omega) -G_{d,-\sigma}^{(0),\adv}(\omega) )
 - G_{d,\sigma}^{(0),\ret}(\omega) (G_{d,\sigma}^{(0),\ret}(\omega)-G_{d,-\sigma}^{(0),\ret}(\omega))] 
\nnr &
 -i(f_{d+}-f_{d-}) [G_{d,+}^{(0),\ret}(\omega) G_{d,-}^{(0),\adv}(\omega) - G_{d,-}^{(0),\ret}(\omega) G_{d,+}^{(0),\adv}(\omega) ] \biggr] . \label{tildemuone}
\end{align}

The spin current in the laboratory frame is therefore 
\begin{align}
\jv_{\rm s} 
=& -\frac{1}{2} \sum_{\kv} (\hop_{\kv})^2 \sumom 
  \rho_\kv(\omega) \lt[
 (\nv\times\dot{\nv})_\alpha {\tilde{\mu}}_1
 + \dot{\nv}_\alpha  \tilde{\mu}_2 \rt].
\label{Jsresult}
\end{align}
In the asymptotic regime, contributions containing only retarded or advanced Green's functions disappear owing to  rapid oscillation, leaving only the contributions containing both retarded and advanced Green's functions.
This asymptotic result agrees with the one based on the pumping formula in the potential scattering picture (Eq. (\ref{jresultmu})).

%

\end{document}